\documentclass[a4paper]{article}%
\usepackage[latin1]{inputenc}
\usepackage{amsmath}%
\usepackage{amsfonts}%
\usepackage{amssymb}%
\usepackage{graphicx}

\begin{document}

\title{Introduction to Nonextensive Statistical Mechanics and Thermodynamics \thanks{To appear in the Proceedings of  the 1953-2003 Jubilee ``Enrico Fermi" International Summer School of Physics {\it The Physics of Complex Systems: New Advances \& Perspectives}, Directors F. Mallamace and H.E. Stanley (1-11 July 2003, Varenna sul lago di Como). The present manuscript reports the content of the lecture delivered by C. Tsallis, and is based on the corresponding notes prepared by F. Baldovin, R. Cerbino  and P. Pierobon, students at the School.}}
\author{Constantino Tsallis$^a$, Fulvio Baldovin$^a$, Roberto Cerbino$^b$ \\and Paolo Pierobon$^{c,d}$ \thanks{tsallis@cbpf.br, baldovin@cbpf.br, Roberto.Cerbino@fisica.unimi.it, pierobon@hmi.de}\\ \\
{\it $^a$Centro Brasileiro de Pesquisas Fisicas}\\ 
{\it Xavier Sigaud 150, 22290 - 180 Rio de Janeiro - RJ, 
Brazil}\\ \\
{\it $^b$Istituto Nazionale per la Fisica della Materia} \\{\it Dipartimento di Fisica, 
Università degli Studi di Milano}\\ {\it Via Celoria 16, 20133 Milano, Italy}\\ \\
{\it $^c$Hahn-Meitner Institut, Abteilung Theorie}\\ {\it Glienicker Strasse 100,
D-14109 Berlin, Germany}\\ \\
{\it $^d$Fachbereich Physik, Freie Universität Berlin} \\{\it Arnimallee 14, 14195
Berlin, Germany}}

\maketitle

\begin{abstract}
In this lecture we briefly review the definition, consequences and applications of an entropy, $S_q$, which generalizes
the usual Boltzmann-Gibbs entropy $S_{BG}$ ($S_1=S_{BG}$), basis of the usual statistical mechanics, well known to be applicable whenever ergodicity is satisfied at the microscopic dynamical level. Such  entropy $S_q$ is based on the notion of $q$-exponential and presents properties not shared by
other available alternative generalizations of $S_{BG}$. The
thermodynamics proposed in this way is generically {\it nonextensive} in a sense that will be qualified. The present framework
seems to describe quite well a vast class of natural and artificial systems which are not ergodic nor close to it.  The a priori calculation of $q$ is
necessary to complete the theory and we present some models where this has already been achieved.

\end{abstract}

\section{Introduction}

Entropy emerges as a classical thermodynamical concept in the 19th
century with Clausius but it is only due to the work of Boltzmann and Gibbs
that the idea of entropy becomes a cornerstone of statistical
mechanics. As result we have that the entropy $S$ of a system is given by the so
called Boltzmann-Gibbs (BG) entropy%
\begin{equation}
S_{BG}=-k\sum_{i=1}^{W}p_{i}\ln p_{i}\label{boltzmann}%
\end{equation}
with the normalization condition%
\begin{equation}
\sum_{i=1}^{W}p_{i}=1\label{normalization} \;.
\end{equation}
Here $p_{i}$ is the probability for the system to be in the $i$-th microstate, and $k$ is an arbitrary constant that, in the framework of
thermodynamics, is taken to be the Boltzmann constant ($k_{B}=1.38\times
10^{-23}$ J/K). Without loss of generality one can also arbitrarily assume $k=1$.
If every microstate has the same probability $p_{i}=1/W$ ({\it equiprobability}
assumption) one obtains the famous {\it Boltzmann principle}
\begin{equation}
S_{BG}=k\ln W\label{boltzmannequi}\;.
\end{equation}

It can be easily shown that entropy (\ref{boltzmann}) is {\it nonnegative}, {\it concave}, {\it extensive} and {\it stable}  \cite{lesche} (or {\it experimentally robust}). By extensive we mean the fact that, if $A$ and $B$ are two {\it independent} systems in the sense that $p_{ij}^{A+B}=p_i^Ap_j^B$, then we straightforwardly verify that
\begin{equation}
S_{BG}(A+B)=S_{BG}(A)+S_{BG}(B) \label{extensive} \;.
\end{equation}
Stability will be addressed later on. One might naturally expect that the form  (\ref{boltzmann}) of $S_{BG}$ would be rigorously {\it derived} from microscopic dynamics. However, the difficulty of performing such a program can be seen from the fact that still today this has not yet been accomplished from first principles. Consequently
(\ref{boltzmann})  is in practice a {\it postulate}. To better realize this point, let us place it on some historical background.

Albert Einstein  says in 1910 \cite{einstein}: \\
`` In order to calculate W, one
needs a \textit{complete} (molecular-mechanical) theory of the system under
consideration. Therefore it is dubious whether the Boltzmann principle has any
meaning without a \textit{complete} molecular mechanical theory or some other
theory which describes the elementary processes. $S=k\ln W$ + {\it constant} seems
without content, from a phenomenological point of view, without giving in
addition such an {\it Elementartheorie}.''. 

In his famous book {\it Thermodynamics}, 
Enrico Fermi  says in 1936 \cite{fermi}:\\
``The entropy of a system composed of several parts is very often equal to the sum
of the entropies of all the parts. This is true if the energy of the system is
the sum of the energies of all the parts and if the work performed by the
system during a transformation is equal to the sum of the amounts of work
performed by all the parts. Notice that these conditions are not quite obvious
and that in some cases they may not be fulfilled.''. 

Laszlo Tisza  says in 1961 \cite{tisza}: \\
``The situation is different for the additivity 
postulate $P\; a2$, the validity of which cannot be inferred from general 
principles. We have to require that the interaction energy between 
thermodynamic systems be negligible. This assumption is closely related 
to the homogeneity postulate $P\; d1$. From the molecular point of view, 
additivity and homogeneity can be expected to be reasonable 
approximations for systems containing many particles, provided that the 
intramolecular forces have a short range character.".

Peter Landsberg  says in 1978 \cite{landsberg}: \\
``The presence of long-range 
forces causes important amendments to thermodynamics, some of which are 
not fully investigated as yet.".

If we put all this together, as well as many other similar statements available in the literature, we may conclude that physical entropies different from the BG one could exist which would be the appropriate ones for anomalous systems. Among the anomalies that we may focus on we include (i) metaequilibrium (metastable) states in large systems involving long range forces between particles, (ii) metaequilibrium states in small systems, i.e., whose number of particles is relatively small, say up to 100-200 particles, (iii) glassy systems, (iv) some classes of dissipative systems, (v) mesoscopic systems with nonmarkovian memory, and others which, in one way or another, might violate the usual simple ergodicity. Such systems might have a multifractal, scale-free or hierarchical structure in their phase space. 

In this spirit, an entropy, $S_q$, which generalizes $S_{BG}$, has been proposed in 1988 \cite{tsallis} as the basis for generalizing BG statistical mechanics. The entropy $S_q$  (with $S_1=S_{BG}$) depends on the index $q$, a real number to be determined {\it a priori} from the microscopic dynamics. This entropy 
seems to describe quite well a large number of natural and artificial systems. As we shall see, the property chosen to be generalized is extensivity, i.e., Eq. (\ref{extensive}). In this lecture we will introduce, through a {\it metaphor}, the form of $S_q$, and will then describe its properties and applications as they have emerged during the last 15 years.

A clarification might be worthy. Why introducing $S_q$ through a metaphor, why not {\it deducing} it? If we knew how to deduce $S_{BG}$ from first principles for those systems (e.g., short-range-interacting Hamiltonian systems) whose microscopic dynamics ultimately leads to ergodicity, we could try to generalize along that path. But this procedure is still unknown, the form (\ref{boltzmann}) being adopted, as we already mentioned, at the level of a {\it postulate}. It is clear that we are then obliged to do the same for any generalization of it. Indeed, there is no logical/deductive way to generalize any set of postulates that are useful for theoretical physics. The only way to do that is precisely through some kind of metaphor.

A statement through which we can measure the difficulty of (rigorously) making the main features of BG statistical mechanics to descend from (nonlinear) dynamics is that of the mathematician Floris Takens. He said in 1991 \cite{takens}: \\
 ``The values of $p_i$ are determined by the following 
dogma: if the energy of the system in the $i^{th}$ state is $E_i$ and 
if the temperature of the system is $T$ then: 
$p_i=\exp\{-E_i/kT\}/Z(T)$, where $Z(T)=\sum_i \exp\{-E_i/kT\}$, (this 
last constant is taken so that $\sum_ip_i=1$). This choice of $p_i$ is called {\it Gibbs distribution}. We shall 
give no justification for this dogma; even a physicist like Ruelle 
disposes of this question as ``deep and incompletely clarified". "  \\
It is a tradition in mathematics to use the word ``dogma" when no theorem is available. Perplexing as it might be for some readers, no theorem is available which establishes, on detailed microscopic dynamical grounds, the necessary and sufficient conditions for being valid the use of the celebrated BG factor. We may say that, at the bottom line, this factor is ubiquitously used in theoretical sciences because seemingly ``it works" extremely well in an enormous amount of cases. It just happens that more and more systems (basically the so called {\it complex systems}) are being identified nowadays where that statistical factor seems to {\it not} ``work"!

\section{Mathematical properties}

\subsection{A metaphor}

The simplest ordinary differential equation one might think of is
\begin{equation}
\frac{dy}{dx}=0\label{odconst} \;,
\end{equation}
whose solution (with initial condition $y(0)=1$) is
$y=1$. 
The next simplest differential equation might be thought to be
\begin{equation}
\frac{dy}{dx}=1\label{odretta}\;,
\end{equation}
whose solution, with the same initial condition, is 
$y=1+x$.
The next one in increasing complexity that we might wish to consider is
\begin{equation}
\frac{dy}{dx}=y\label{odexp}\;,
\end{equation}
whose solution is $y=e^{x}$. Its inverse function is
\begin{equation}
y=\ln x\label{log}\;,
\end{equation}
which has the same functional form of the Boltzmann-Gibbs entropy
(\ref{boltzmannequi}), and satisfies the well known additivity property%
\begin{equation}
\ln(x_{A}x_{B})=\ln x_{A}+\ln x_{B}.\label{additivity}%
\end{equation}
A question that might be put is: can we unify all three cases  (\ref{odconst},\ref{odretta},\ref{odexp}) considered above? A trivial positive answer would be to consider $dy/dx =a+by$, and play with $(a,b)$. Can we unify with {\it only one} parameter? The answer still is positive, but this time {\it out of linearity}, namely with
\begin{equation}
\frac{dy}{dx}=y^{q} \;\;\;(q \in {\cal R}) \;,              \label{odqexp}%
\end{equation}
which, for $q\rightarrow -\infty, \;q=0$ and $q=1$, reproduces respectively the 
differential equations (\ref{odconst}), (\ref{odretta}) and (\ref{odexp}).
The solution of (\ref{odqexp}) is given by the {\it $q$-exponential} function
\begin{equation}
y=[1+(1-q)x]^{\frac{1}{1-q}}\equiv e_{q}^{x}\ \ \ \ \ \ (e_{1}^{x} 
=e^{x})  \;,\label{qexp}%
\end{equation}
whose inverse is the {\it $q$-logarithm} function
\begin{equation}
y=\frac{x^{1-q}-1}{1-q}\equiv\ln_{q}x\text{ \ \ \ \ \ \ }(\ln_{1}x=\ln
x).\label{qln}%
\end{equation}
This function satisfies the {\it pseudo-additivity} property
\begin{equation}
\ln_{q}(x_{A}x_{B})=\ln_{q}x_{A}+\ln_{q}x_{B}+(1-q)(\ln_{q}x_{A})(\ln_{q}%
x_{B})\label{pseudoadditivity}%
\end{equation}

\subsection{The nonextensive entropy $S_q$}

We can rewrite Eq. (\ref{boltzmann}) in a slightly different form, namely (with $k=1$)
\begin{equation}
S_{BG}=-\sum_{i=1}^{W}p_{i}\ln p_{i}=\sum_{i=1}^{W}p_{i}\ln\frac{1}{p_{i}%
}=\left\langle \ln\frac{1}{p_{i}}\right\rangle     \;,\label{boltzmannsurprise}%
\end{equation}
where $\left\langle ...\right\rangle \equiv\sum_{i=1}^{W}(...)p_{i}$. The 
quantity $ln(1/p_{i})$ is sometimes called \textit{surprise} or \textit{unexpectedness}. Indeed, 
$p_{i}=1$ corresponds to certainty, hence zero surprise if the {\it expected} event does occur; on the other hand, $p_{i} \to 0$ corresponds to nearly impossibility, hence
infinite surprise if the {\it unexpected} event does occur. If we introduce the {\it $q$-surprise} (or {\it $q$-unexpectedness}) as $ln_{q}(1/p_{i})$, it is
kind of natural to define the following $q$-entropy 
\begin{equation}
S_{q}\equiv \left\langle \ln_{q}\frac{1}{p_{i}}\right\rangle =\sum_{i=1}^{W}p_{i}%
\ln_{q}\frac{1}{p_{i}}=\frac{1-\sum_{i=1}^{W}p_{i}^{q}}{q-1}%
\label{tsallissurprise}%
\end{equation}
In the limit $q\rightarrow1$ one has $p_{i}^{q}=pe^{(q-1)\ln p_{i}}\sim
p_{i}[1+(q-1)\ln p_{i}]$, and the entropy $S_q$ coincides with the
Boltzmann-Gibbs one, i.e., $S_1=S_{BG}$.
Assuming equiprobability (i.e., $p_{i}=1/W$) one obtains straightforwardly
\begin{equation}
S=\frac{W^{1-q}-1}{1-q}=\ln_{q}W.\label{qboltzmann}%
\end{equation}

Consequently, it is clear that $S_q$ is a generalization of and not an alternative to
the Boltzmann-Gibbs entropy. The pseudo-additivity of the $q$-logarithm immediately implies 
(for the following formula we restore arbitrary $k$)
\begin{equation}
\frac{S_{q}(A+B)}{k}=\frac{S_{q}(A)}{k}+\frac{S_{q}(B)}{k}+(1-q)\frac{S_{q}%
(A)}{k}\frac{S_{q}(B)}{k}\label{qsumentropy}%
\end{equation}
if A\ and B are two {\it independent} systems (i.e., $p_{ij}^{A+B}=p_{i}^{A}p_{j}^{B}$). It follows that $q=1$, $q<1$ and $q>1$ respectively correspond to the {\it extensive}, {\it superextensive} and {\it subextensive} cases. It is from this property that the corresponding generalization of the BG statistical mechanics is often referred to as
\textit{nonextensive statistical mechanics}. 

Eq. (\ref{qsumentropy}) is true under the hypothesis of independency between A and B.  But if they are correlated in some special, strong way, it may
exist $q^{\ast}$ such that
\begin{equation}
S_{q^{\ast}}(A+B)=S_{q^{\ast}}(A)+S_{q^{\ast}}(B) \label{sumentropy}\;, 
\end{equation}
thus recovering extensivity, but of a {\it different} entropy, not the usual one!
Let us illustrate this interesting point through two examples:


(i) A system of $N$ nearly independent elements yields
$W(N)\sim\mu^{N}$ (with $\mu>1$) (e.g., $\mu=2$ for a coin, $\mu=6$ for a dice). Its entropy $S_q$ is
given by
\begin{equation}
S_{q}(N)=\ln_{q}W(N)\sim\frac{\mu^{N(1-q)}-1}{1-q}%
\end{equation}
and extensivity is obtained {\it if and only if} $q=1$. In other words, 
$S_{1}(N)\sim N\ln\mu\varpropto N$. This is the usual case, discussed in any textbook.

(ii) A system whose elements are correlated at all scales might correspond to
$W(N)\sim N^{\rho}$ (with $\rho>0$). Its entropy $S_q$ is given
by
\begin{equation}
S_{q}(N)=\ln_{q}W(N)\sim\frac{N^{\rho(1-q)}-1}{1-q}%
\end{equation}
and extensivity is obtained {\it if and only if} $q=q^{\ast} \equiv 1-\frac{1}{\rho}<1$ . In other words, 
$S_{q^{\ast}}(N)\varpropto N$. It is allowed to think that such possibility could please as much as the usual one (present example (i)) somebody --- like Clausius --- wearing the ``glasses" of classical thermodynamics! Indeed, it does not seem that Clausius had in mind any specific functional form for the concept of entropy he introduced, but he was surely expecting it to be proportional to $N$ for large $N$. We see, in both examples that we have just analyzed, that it can be so. In particular, the entropy $S_q$ is nonextensive for independent systems, but {\it can be extensive for systems whose elements are highly correlated}. 

\subsection{The 1988 derivation of the entropy $S_q$}

Let us point out that the 1988 derivation \cite{tsallis}\ of the entropy $S_q$ was slightly different. 
If we consider $0<p_{i}<1$\ and $q>0$ we have that $p_{i}^{q}\gtrless p_{i}$
if $q\lessgtr1$ and $p_{i}^{q}=p_{i}$ if $q=1$. Many physical phenomena seem to be controlled either by the rare events or the common events. Similarly to what occurs in multifractals, one could take such effect into account by choosing $q<1$ (hence {\it low probabilities are enhanced and high probabilities are depressed}; indeed, $\lim_{p_i \to 0}p_i^q/p_i \to \infty$) or $q>1$ (hence {\it the other way around}; indeed, $\lim_{p_i \to 0}p_i^q/p_i =0$).  The present index $q$ is completely different from the multifractal index (currently noted $q_M$ in the nonextensive statistical literature to avoid confusion, where $M$ stands for {\it multifractal}), but the type of bias is obviously analogous. What happens then if we introduce this type of \textit{bias} into the entropy itself?

At variance with the concept of energy, the entropy should {\it not} depend on the physical support of the information, hence it must be invariant under permutations of the events. The most natural choice is then to make it depend on $\sum_{i=1}^{W}p_{i}^{q}$. The simplest dependence being the linear one, one might propose $S_q=A+B\sum_{i=1}^{W}p_{i}^{q}$. If some event has $p_{i}=1$ we have certainty, and we consequently expect $S_{q}=0$, hence $A+B=0$, hence
$S(\{p_{i}\})=A(1-\sum_{i=1}^{W}p_{i}^{q})$.
We impose now $S_{1}=S_{BG}$ and, taking into account $p_{i}^{q}=p_{i}e^{(q-1)\ln p_{i}}\sim p_{i}[1+(q-1)\ln p_{i}]$,  we obtain 
$S_{1}\sim-A(q-1)\sum_{i=1}^{W}p_{i}\ln p_{i}$. 
So we choose $A=\frac{1}{q-1}$
and straightforwardly reobtain (\ref{tsallissurprise}).

\subsection{$S_{q}$ and the Shannon-Khinchin axioms}

Shannon in 1948 and then Khinchin in 1953 gave quite similar sets of axioms about the form of the entropic
functional (\cite{shannon},\cite{khinchin}). Under reasonable requests about entropy they obtained that the only functional form allowed is the Boltzmann-Gibbs entropy.

\paragraph{Shannon theorem:}
\begin{enumerate}
\item $S(p_{1},...,p_{W})$ continuous function with respect to all its arguments

\item $S(p_{1}=p_2=...=p_W=\frac{1}{W})$ monotonically increases with W

\item $S(A+B)=S(A)+S(B)$ $\ \ \ \ $if$\ \ p_{ij}^{A+B}=p_{i}^{A}p_{j}^{B}$

\item $S(p_{1},...,p_{W})=S(p_{L},p_{M})+p_{L}S(\frac{p_{1}}{p_{L}%
},...,\frac{p_{W_L}}{p_{L}})+p_{M}S(\frac{p_{W_L+1}}{p_{M}},...,\frac{p_{W}}{p_{M}%
})$ where $W=W_{L}+W_{M}$, $p_{L}=\sum_{i=1}^{W_{L}}p_{i}$, $p_{M}%
=\sum_{i=W_{L}+1}^{W}p_{i}$ (hence $p_{L}+p_{M}=1$)
\end{enumerate}

{\it if and only if} $S(p_{1},...,p_{W})$ is given by Eq. (1).

\paragraph{Khinchin theorem:}

\begin{enumerate}
\item $S(p_{1},...,p_{W})$ continuous function with respect to all its arguments

\item $S(p_{1}=p_2=...=p_W=\frac{1}{W})$ monotonically increases with W

\item $S(p_{1},...,p_{W},0)=S(p_{1},...,p_{W})$

\item $S(A+B)=S(A)+S(B|A)$, $S(B|A)$ being the conditional entropy
\end{enumerate}

{\it if and only if} $S(p_{1},...,p_{W})$ is given by Eq. (1).\\

The following generalizations of these theorems have been given by Santos in 1997 
\cite{santos} and by Abe in 2000 \cite{abe}. The latter was first conjectured by Plastino and Plastino in 1996 \cite{plastinosCondMat} and 1999 \cite{plastinosBJP}.

\paragraph{Santos theorem:}

\begin{enumerate}
\item $S(p_{1},...,p_{W})$ continuous function with respect to all its arguments

\item $S(p_{1}=p_2=...=p_W=\frac{1}{W})$ monotonically increases with W

\item $\frac{S(A+B)}{k}=\frac{S(A)}{k}+\frac{S(B)}{k}+(1-q)\frac{S_{q}(A)}%
{k}\frac{S_{q}(B)}{k}$ $\ \ \ \ $if$\ \ p_{ij}^{A+B}=p_{i}^{A}p_{j}^{B}$

\item $S(p_{1},...,p_{W})=S(p_{L},p_{M})+p_{L}^qS(\frac{p_{1}}{p_{L}%
},...,\frac{p_{W_L}}{p_{L}})+p_{M}^qS(\frac{p_{W_L+1}}{p_{M}},...,\frac{p_{W}}{p_{M}%
})$ where $W=W_{L}+W_{M}$, $p_{L}=\sum_{i=1}^{W_{L}}p_{i}$, $p_{M}%
=\sum_{i=W_{L}+1}^{W}p_{i}$ (hence $p_{L}+p_{M}=1$)
\end{enumerate}

{\it if and only if}
\begin{equation}
S(p_{1},...,p_{W})=k\frac{1-\sum_{i=1}^{W}p_{i}^{q}}{q-1}%
\end{equation}
\paragraph{Abe theorem:}

\begin{enumerate}
\item $S(p_{1},...,p_{W})$ continuous function with respect to all its arguments

\item $S(p_{1}=p_2=...=p_W=\frac{1}{W})$ monotonically increases with W

\item $S(p_{1},...,p_{W},0)=S(p_{1},...,p_{W})$

\item $\frac{S(A+B)}{k}=\frac{S(A)}{k}+\frac{S(B|A)}{k}+(1-q)\frac{S_{q}%
(A)}{k}\frac{S(B|A)}{k}$, $S(B|A)$ being the conditional entropy
\end{enumerate}

{\it if and only if} 
$S(p_{1},...,p_{W})$ is given by Eq. (21).\\

The Santos and the Abe theorems clearly are important. Indeed, they show that the entropy $S_q$ is the {\it only}
possible entropy that extends the Boltzmann-Gibbs entropy maintaining the basic 
properties but allowing, if $q\neq1$, nonextensivity (of the form of Santos' third axiom, or Abe's fourth axiom). 

\subsection{Other mathematical properties}

\paragraph{Reaction under bias:}

It has been shown \cite{abe2}\ that the Boltzmann-Gibbs entropy can be
rewritten as%
\begin{equation}
S_{BG}=-\left[  \frac{d}{dx}\sum_{i=1}^{W}p_{i}^{x}\right]  _{x=1} \;.%
\end{equation}
This can be seen as a reaction to a translation of the bias $x$ in the same way as
differentiation can be seen as a reaction of a function under a (small) {\it translation} of the abscissa.
Along the same line, the entropy $S_q$ can be rewritten as
\begin{equation}
S_{q}=-\left[  D_{q}\sum_{i=1}^{W}p_{i}^{x}\right]  _{x=1}\;,
\end{equation}
where
\begin{equation}
D_{q}h(x)\equiv\frac{h(qx)-h(x)}{qx-x} \;\;\;\;\;\Bigl(D_{1}h(x)=\frac{dh(x)}{dx}\Bigr)
\end{equation}
is the Jackson's 1909 generalized derivative, which can be seen as a reaction of a function under
{\it dilatation} of the abscissa (or under a {\it finite} increment of the abscissa).

\paragraph{Concavity:}
If we consider two probability distributions $\{p_{i}\}$ and $\{p_{i}^{\prime }\}$\ for a
given system $(i=1,...,W)$, we can define the {\it convex sum} of the
two probability distributions as%
\begin{equation}
p_{i}^{\prime\prime}\equiv\mu p_{i}+(1-\mu)p_{i}^{\prime}\text{
\ \ \ \ \ \ \ \ \ \ (0$<$ }\mu\text{$<$1).}
\end{equation}
An entropic functional $S(\{p_{i}\})$ is said \textit{concave} if and only if
for all $\mu$ and for all $\{p_{i}\}$ and $\{p_{i}^{\prime}\}$%
\begin{equation}
S(\{p_{i}^{\prime\prime}\})\geq\mu S(\{p_{i}\})+(1-\mu)S(\{p_{i}^{\prime}\})
\end{equation}
By convexity we mean the same property where $\geq$ is replaced by $\leq$. It can be
easily shown that the entropy $S_q$ is concave (convex) for every $\{p_i\}$ and every $q>0$
($q<0$). It is important to stress that this property implies, in the framework of
statistical mechanics, thermodynamic stability, i.e., stability of the system with regard
to energetic perturbations. This means that the entropic functional is defined such that
the stationary state (e.g., thermodynamic equilibrium) makes it extreme (in the present case,
maximum for $q>0$ and minimum for $q<0$). Any perturbation of $\{p_i\}$ which makes the
entropy extreme is followed by a tendency toward $\{p_i\}$ once again. Moreover, such a
property makes possible for two systems at different temperature to equilibrate to a
common temperature.

\paragraph{Stability or experimental robustness:}
An entropic functional $S(\{p_{i}\})$ is said {\it stable} or \textit{experimentally robust} if and only
if, for any given $\varepsilon>0$, exists $\delta
_{\varepsilon}>0$ such that, independently from $W$,
\begin{equation}
\sum_{i=1}^{W}\left|  p_{i}-p_{i}^{\prime}\right|  \leq\delta_{\varepsilon
}\Rightarrow\left|  \frac{S(\{p_{i}\})-S(\{p_{i}^{\prime}\})}{S_{\max}%
}\right|  <\varepsilon \;.
\end{equation}
This implies in particular that
\begin{equation}
\lim_{\varepsilon\rightarrow0}\lim_{W\rightarrow \infty}\left|  \frac{S(\{p_{i}%
\})-S(\{p_{i}^{\prime}\})}{S_{\max}}\right|  =\lim_{W\rightarrow \infty}%
\lim_{\varepsilon\rightarrow0}\left|  \frac{S(\{p_{i}\})-S(\{p_{i}^{\prime
}\})}{S_{\max}}\right|  =0 \;.
\end{equation}

Lesche \cite{lesche} has argued that experimental robustness is a necessary requisite for
an entropic functional to be a physical quantity because essentially assures that, under
arbitrary small variations of the probabilities, the relative variation of entropy remains small.
This property is to be not confused with {\it thermodynamical stability}, considered above. It has been shown \cite{abe3} that the entropy $S_q$  exhibits, for any $q>0$, 
this property.

\subsection{A remark on other possible generalizations of the \\ Boltzmann-Gibbs entropy}

There have been in the past other generalizations of the BG entropy. The {\it Renyi
entropy} is one of them and is defined as follows
\begin{equation}
S_{q}^{R} \equiv \frac{\ln\sum_{i=1}^{W}p_{i}^{q}}{1-q}=\frac{\ln[1+(1-q)S_{q}]}%
{1-q} \;.
\end{equation}

Another entropy has been introduced by Landsberg and Vedral \cite{lv} and independently by
Rajagopal and Abe \cite{ra}. It is sometimes called {\it normalized nonextensive entropy}, and is defined as follows
\begin{equation}
S_q^N \equiv S_{q}^{LVRA}\equiv \frac{1-\frac{1}{\sum_{i=1}^{W}p_{i}^{q}}}{1-q}=\frac{S_{q} %
}{1+(1-q)S_{q}} \;.
\end{equation}

A question arises naturally: Why not using one of these entropies (or even a different one such as the so called {\it escort entropy} $S_q^E$, defined in \cite{tsallis98,brigatti}), instead of $S_q$, for generalizing BG statistical mechanics? The answer appears to be quite straightforward.
$S_{q}^{R}$, $S_{q}^{LVRA}$ and $S_q^E$ are not concave nor experimentally robust. Neither yield they a {\it finite} entropy production for unit time, in contrast with $S_q$, as we shall see later on. Moreover, these alternatives do not possess the suggestive structure that $S_q$ exhibits associated with the Jackson generalized derivative.  
Consequently, {\it for thermodynamical purposes}, it seems nowadays quite natural to consider the entropy $S_q$ as the best candidate for generalizing the Boltzmann-Gibbs entropy. It might be different for other purposes: for example, Renyi entropy is known to be useful for geometrically characterizing multifractals. 

\section{Connection to thermodynamics}
Dozens of entropic forms have been proposed during the last decades, but not all of them are necessarily related to the physics of nature. 
Statistical mechanics is more than the adoption of an entropy: the (meta)equilibrium 
probability distribution must not only optimize the entropy but also satisfy, in addition to the norm constraint,  
constraints on quantities such as the energy. 
Unfortunately, when the theoretical frame is generalized, it is not obvious which constraints are to be maintained and which ones are to be generalized, and in what manner. In this section we derive, following along the lines of Gibbs, a thermodynamics for (meta)equilibrium distribution based on the entropy defined above. 
It should be stressed that the distribution derived in this way for $q\neq1$ does {\it not} correspond to thermal equilibrium (as addressed within the BG formalism through the celebrated Boltzmann's \emph{molecular chaos hypothesis}) but rather to a metaequilibrium or a stationary state, suitable to describe a large class of nonergodic systems.

\subsection{Canonical ensemble}
For a system in thermal contact with a large reservoir, and in analogy with the path followed by Gibbs \cite{tsallis,tsallis98}, we look for the 
distribution which optimizes the entropy $S_q$ defined in Eq. (21), 
with the normalization condition (2) and the following constraint on the energy \cite{tsallis98}:
\begin{equation}\label{con2}
\frac{\sum_{i=1}^{W}p_i^qE_i}{\sum_{j=1}^{W}p_j^q}=U_q
\end{equation}
where
\begin{equation}\label{escort}
P_i\equiv\frac{p_i^q}{\sum_{j=1}^Wp_j^q}
\end{equation}
is called \emph{escort distribution}, and $\{E_i\}$ are the eigenvalues of the system Hamiltonian with the chosen boundary conditions. Note that, in analogy with BG statistics, a constraint
like $\sum_{i=1}^Wp_iE_i=U$ would be more intuitive. This was indeed the first attempt
\cite{tsallis}, but though it correctly yields, as stationary (meta-equilibrium) distribution, 
the $q$-exponential, it turns out to be inadequate for various reasons, including related to 
Lévy-like superdiffusion, for which a diverging second moment exists. 

Another natural choice \cite{tsallis,curado} would be to fix
$\sum_{i=1}^Wp_i^qE_i$ but, though this solves annoying divergences, it creates some new
problems: the metaequilibrium distribution is {\it not} invariant under change of the zero level
for the energy scale, $\sum_{i=1}^W p_i^q\neq1$ implies that the constraint applied to a
constant does {\it not} yield the same constant, and above all, the assumption
$p_{ij}^{A+B}=p_i^Ap_j^B$ and $E_{ij}^{A+B}=E_i^A+E_j^B$ does {\it not} yield
$U_{ij}^{A+B}=U_i^A+U_j^B$, i.e., the energy conservation principle is {\it not} the same in the microscopic and macroscopic worlds. 

It is by now well established that the energy constraint must be imposed in the form (\ref{con2}), using the normalized escort
distribution (\ref{escort}). A detailed discussion of this important point can be found in \cite{tsallisinswinney}.

The optimization of $S_q$ with the constraints
(2) and (\ref{con2}) with Lagrange multiplier $\beta$ yields:
\begin{equation}
p_i=\frac{\left[1-(1-q)\beta_q(E_i-U_q)\right]^{\frac{1}{1-q}}}{\overline{Z}_q}=\frac{e_q^{-\beta_q(E_i-U_q)}}{\overline{Z}_q}
\end{equation}
with
\begin{equation}
\overline{Z}_q\equiv\sum_{j=1}^We_q^{-\beta_q(E_i-U_q)}
\end{equation}
and
\begin{equation}
\beta_q\equiv\frac{\beta}{\sum_{j=1}^Wp_j^q}
\end{equation}
It turns out that the metaequilibrium distribution can be written hidding the presence of
$U_q$ in a form which sometimes is more convenient when we want to use experimental or
computational data:
\begin{equation}
p_i=\frac{\left[1-(1-q)\beta_q'E_i\right]^{\frac{1}{1-q}}}{Z_q'}=\frac{e_q^{-\beta_q'E_i}}{Z_q'}
\end{equation}
with
\begin{equation}
Z_q'\equiv\sum_{j=1}^We_q^{-\beta_q'E_i}
\end{equation}
and
\begin{equation}
\beta_q'\equiv\frac{\beta}{\sum_{j=1}^Wp_j^q+(1-q)\beta U_q}
\end{equation}
It can be easily checked that 
(i) for $q\to1$, the BG weight is recovered, i.e.,  $p_i=e^{-\beta E_i}/Z_1$, 
(ii) for $q>1$, a power-law tail emerges, and 
(iii) for $q<1$, the formalism imposes a high energy cutoff ($p_i=0$) whenever the
  argument of the $q$-exponential function becomes negative. 

Note that distribution (33) is generically not an exponential law, i.e., it is generically
{\it not} factorizable (under sum in the argument), and nevertheless is invariant under choice
of the zero energy for the energy spectrum (this is one of the pleasant facts associated with the choice of energy constraint in terms of a
normalized distribution like the escort one).

\subsection{Legendre structure}
The Legendre-transformation structure of thermodynamics holds for every $q$ (i.e., it is
$q$-invariant) and allows us to connect the theory developed so far to thermodynamics.

We verify that, for all values $q$, 
\begin{equation}
 \frac{1}{T}=\frac{\partial S_q}{\partial U_q}\;\;\;\;\;\; (T \equiv 1/k\beta)
\end{equation}
Also, it can be proved that the free energy is given by
  \begin{equation}
F_q\equiv U_q-TS_q=-\frac{1}{\beta}\ln_q Z_q\hspace{1cm}
  \end{equation}
where
  \begin{equation}
\ln_qZ_q\equiv\frac{Z_q^{1-q}-1}{1-q}=\frac{\overline{Z}_q^{1-q}-1}{1-q}-\beta U_q \;,
\end{equation}
and the internal energy is given by
\begin{equation}
    U_q=-\frac{\partial}{\partial\beta}\ln_qZ_q \;.
\end{equation}
Finally, the specific heat reads
\begin{equation}
C_q\equiv T\frac{\partial S_q}{\partial T}=\frac{\partial U_q}{\partial 
T}=-T\frac{\partial^2F_q}{\partial T^2} \;.
\end{equation}

In addition to the Legendre structure, many other theorems and properties are $q$-invariant, thus supporting
the thesis that this is a right road for generalizing the BG theory. Let us briefly list some of them.
\begin{enumerate}
  \item Boltzmann H-theorem ({\it macroscopic time irreversibility}): 
  \begin{equation}
q\frac{dS_q}{dt}\geq 0 \;\;\;\;(\forall q)\;.
  \end{equation}
This inequality has been established under a variety of irreversible time evolution mesoscopic equations (\cite{mariz,ramshaw} and others), and is consistent with the second principle of thermodynamics (\cite{secondprinciple}), which turns out to be satisfied for all values of $q$. 
  \item Ehrenfest theorem ({\it correspondence principle between classical and quantum
  mechanics})  : Given an observable $\widehat{{\cal O}}$ and the Hamiltonian $\widehat{{\cal H}}$ of the system, it can be shown (see \cite{plastino93} for unnormalized $q$-expectation values; for the normalized ones, the proof should follow along the same lines) that
\begin{equation}
\frac{d\langle\widehat{{\cal O}}\rangle_q}{dt}=\frac{i}{\hbar}\langle[\widehat{{\cal H}},\widehat{{\cal O}}]\rangle_q  \;\;\;\;(\forall
q) \;.
\end{equation}
  \item Factorization of the likelihood function ({\it thermodynamically independent systems}): The
  likelihood function satisfies \cite{cacerestsallis,chame2,tsallis95}
\begin{equation} 
W_q(\{p_i\})\propto e_q^{S_q(\{p_i\})} \;.
\end{equation} 
Consequently, if $A$ and $B$
  are two probabilistically independent systems, it can be verified that
  \begin{equation}
W_q(A+B)=W_q(A)W_q(B) \;\;\;\;(\forall q) \;,
  \end{equation}
as expected by Einstein \cite{einstein}.
  \item Onsager reciprocity theorem ({\it microscopic time reversibility}): It has been shown
  \cite{ra2,caceres,chame}  that the reciprocal linear coefficients satisfy
  \begin{equation}
L_{jk}=L_{kj}\hspace{1cm}(\forall q) \;,
  \end{equation}
thus satisfying the fourth principle of thermodynamics. 
  \item Kramers and Kronig relations ({\it causality}): They have been proved in \cite{ra2} for all values of
  $q$. 
  \item Pesin theorem ({\it connection between sensitivity to the initial conditions and the
  entropy production per unit time}). We can define the {\it $q$-generalized Kolmogorov-Sinai entropy} as
\begin{equation}  
K_q\equiv\lim_{t\to\infty}\lim_{W\to\infty}\lim_{N\to\infty}\frac{\langle S_q\rangle(t)}{t} \;,
\label{kolmogorov}
\end{equation}
where $N$ is the number of initial conditions, $W$ is the number of windows in the partition ({\it fine graining}) we have adopted, and $t$ is (discrete) time. Let us mention that the standard Kolmogorov-Sinai entropy is defined in a slightly different manner in the mathematical theory of nonlinear dynamical systems. See more details in \cite{pesincatania} and references therein.  

The {\it $q$-generalized Lyapunov coefficient} $\lambda_q$ can be defined through the sensitivity to the initial conditions
\begin{equation}
\xi       \equiv \lim_{\Delta x(0)\to0} \Delta
x(t)/\Delta x(0) =e_q^{\lambda_q t} \;
\label{q-lyap}
\end{equation}
where we have focused on a one-dimensional system (basically $x(t+1)=g(x(t))$, $g(x)$ being a nonlinear function, for example that of the logistic map). It was 
conjectured in 1997 \cite{tsallis97}, and recently proved for unimodal maps \cite{pesinfulvioalberto}, that they are related through
\begin{equation}
K_q=\left\{\begin{array}{ll} \lambda_q&\textrm{ if }\lambda_q>0\\ 0&\textrm{otherwise}
\end{array}
\right.
\end{equation}
To be more explicit, we have $K_1=\lambda_1$ if $\lambda_1 \ge 0$ (and $K_1=0$ if $\lambda_1<0$). But if we have $\lambda_1=0$, then we have a special value of $q$ such that $K_q=\lambda_q$ if $\lambda_q \ge 0$ (and $K_q=0$ if $\lambda_q<0$).
\end{enumerate}
Notice that the $q$-invariance of all the above properties is kind of natural. Indeed, their
origins essentially lie in mechanics, and what we have generalized is {\it not} mechanics but only the
concept of information upon it.

\section{Applications}

The ideas related with nonextensive statistical mechanics have received an enormous amount of applications in a variety of disciplines including physics, chemistry, mathematics, geophysics, biology, medicine, economics, informatics, geography, engineering, linguistics and others. For description and details about these, we refer the reader to \cite{tsallisinswinney,tsallisingellmann} as well as to the bibliography in \cite{biblio}.  The a priori determination  (from microscopic or mesoscopic dynamics) of the index $q$ is illustrated for a sensible variety of systems in these references. This point obviously is a very relevant one, since otherwise the present theory would {\it not} be {\it complete}.

In the present brief introduction we shall address only two types of systems, namely a long-range-interacting many-body classical Hamiltonian, and the logistic-like class of one-dimensional dissipative maps. The first system is still under study (i.e., it is only partially understood), but we present it here because it might constitute a direct application of the thermodynamics developed in the Section 3.  The second system is considerably better understood, and illustrates the various concepts which appear to be relevant in the present generalization of the BG ones.

\subsection{Long-range-interacting many-body classical Hamiltonians}

To illustrate this type of system, let us first focus on the inertial $XY$ ferromagnetic model, characterized by the following Hamiltonian \cite{antoniruffo,celiaconstantino}:
\begin{equation}
{\cal H}=\sum_{i=1}^N  \frac{p_i^2}{2} +
  \sum_{i \ne j}  \frac{1-cos(\theta_i -\theta_j)}{r_{ij}^{\;\alpha}}~~~~(\alpha \ge 0),
\end{equation}
where $\theta_i$ is the $i-th$ angle and $p_i$ the 
conjugate variable   representing   the  angular momentum
(or the rotational velocity since, without loss of generality, unit moment of inertia is assumed).  

The summation in the potential is extended to all couples of spins (counted only once) 
and not restricted to first neighbors; for $d=1$, $r_{ij}=1, 2, 3,...$;  for $d=2$, $r_{ij}=1, \sqrt{2}, 2,...$; for $d=3$, $r_{ij}=1, \sqrt{2}, \sqrt{3}, 2, ...$. The first-neighbor coupling constant has been assumed, without loss of generality, to be equal to unity. This model is an inertial version of the well known $XY$ ferromagnet. Although it does not make any relevant difference, we shall assume periodic boundary conditions, the distance to be considered between a given pair of sites being the smallest one through the $2d$ possibilities introduced by the periodicity of the lattice. Notice that the two-body potential term has been written in such a way as to have zero energy for the global fundamental state (corresponding to $p_i=0$, $ \forall i$, and all $\theta_i$ equal among them, and equal to say zero). The $\alpha \to\infty$ limit corresponds to only first-neighbor interactions, whereas the $\alpha=0$ limit corresponds to infinite-range interactions (a typical Mean Field situation, frequently referred to as the HMF model \cite{antoniruffo}). 

The quantity ${\tilde N} \equiv \sum_{i \ne j} r_{ij}^{\;-\alpha}$ corresponds essentially to the potential energy {\it per rotator}. This quantity, 
in the limit $N \to \infty$, converges to a finite value if $\alpha/d >1$, and diverges like $N^{1-\alpha/d}$ if $0 \le \alpha/d<1$ (like $\ln N$ for $\alpha/d=1$). 
In other words, the energy is extensive for $\alpha/d >1$ and nonextensive otherwise. In the extensive case (here  referred to as {\it short range interactions}; also referred to as {\it integrable interactions} in the literature), the thermal equilibrium (stationary state attained in the $t \to\infty$ limit) is known to be the BG one (see \cite{fisheretal}).  The situation is much more subtle in the nonextensive case ({\it long range interactions}). It is this situation that we focus on here. ${\tilde N}$ behaves like $\int_1^{N^{1/d}} dr \; r^{d-1} r^{-\alpha} \propto \frac{N^{1-\alpha/d}-1}{1-\alpha/d} \sim \frac{N^{1-\alpha/d}-\alpha/d}{1-\alpha/d}$. All these three equivalent quantities ($\tilde N$ or $\frac{N^{1-\alpha/d}-1}{1-\alpha/d}$ or $ \frac{N^{1-\alpha/d}-\alpha/d}{1-\alpha/d}$) are indistinctively used in the literature to scale the energy per particle of such long-range systems. In order to conform to the most usual writing, we shall from now on replace the Hamiltonian ${\cal H}$ by the following rescaled one:
\begin{equation}
{\cal H^\prime}=\sum_{i=1}^N  \frac{p_i^2}{2} +
  \frac{1}{{\tilde N}}\sum_{i \ne j}  \frac{1-cos(\theta_i -\theta_j)}{r_{ij}^{\;\alpha}}~~~~(\alpha \ge 0),
\end{equation} 
The molecular dynamical results associated with this Hamiltonian (now artificially transformed into an extensive one for {\it all} values of $\alpha/d$) can be trivially transformed into those associated with Hamiltonian ${\cal H}$ by re-scaling time (see \cite{celiaconstantino}). 

Hamiltonian (53) exhibits in the microcanonical case (isolated system at fixed total energy $U$) a second order phase transition at $u \equiv U/N = 0.75$. It has anomalies both above and below this critical point. 

Above the critical point it has a Lyapunov spectrum which, in the $N \to \infty $ limit, approaches, for $0 \le \alpha/d \le1$, zero as $N^{-\kappa}$, where $\kappa(\alpha/d)$ decreases from $1/3$ to zero when $\alpha/d$ increases from zero to unity, and remains zero for $\alpha/d \ge 1$ \cite{celiaconstantino,giansanti}. It has a Maxwellian distribution of velocities \cite{vitoandreaconstantino}, and exhibits no aging \cite{aging}. Although it has no aging, the typical correlation functions depend on time as a $q$-exponential.  Diffusion is shown to be of the normal type.

Below the critical point (e.g., $u=0.69$), for a nonzero-measure class of initial conditions, a longstanding quasistationary (or metastable) state precedes the arrival to the BG thermal equilibrium state. The duration of this quasistationary state appears to diverge with $N$ like $\tilde N$ \cite{vitoandreaconstantino,cabral}. During this anomalous state, there is aging (the correlation functions being well reproduced by $q$-exponentials once again), and the velocity distribution is not Maxwellian, but rather approaches a $q$-exponential function (with a cutoff at high velocities, as expected for any microcanonical system). Anomalous superdiffusion is shown to exist in this state. The mean kinetic energy ($\propto T$, where $T$ is referred to as the {\it dynamical temperature}) slowly approaches the BG value from below, the relaxation function being once again a $q$-exponential one. During the anomalous aging state, the zeroth principle of thermodynamics and the basic laws of thermometry have been shown to hold as usual \cite{tsallisreply,moyano}. The fact that such basic principles are preserved constitutes a major feature, pointing towards the applicability of thermostatistical arguments and methods to this highly nontrivial quasistationary state. 

Although none of the above indications constitutes a proof that this long-range system obeys, in one way or another, nonextensive statistical mechanics, the set of so many consistent evidences may be considered as a very strong suggestion that so it is. Anyhow, work is in progress to verify closely this tempting possibility (see also \cite{physicaa}).

Similar observations are in progress for the Heisenberg version of the above Hamiltonian \cite{nobre}, as well as for a  $XY$ model including a local term which breaks the angular isotropy in such a way as to make the model to approach the Ising model  \cite{ernestoising}. 

Lennard-Jones small clusters (with $N$ up to 14) have been numerically studied recently \cite{doye}. The distributions of the number of local minima of the potential energy  with $k$ neighboring saddle-points in the configurational phase space can, although not mentioned in the original paper \cite{doye}, be quite well fitted with $q$-exponentials with $q=2$. No explanation is still available for this suggestive fact. Qualitatively speaking, however, the fact that we are talking of very {\it small} clusters makes that, despite the fact that the Lennard-Jones interaction is not a long-range one thermodynamically speaking (since $\alpha/d=6/3>1$), all the atoms sensibly see each other, therefore fulfilling a nonextensive scenario.  

\begin{figure}
\begin{center}
\includegraphics[width=.65\textwidth]{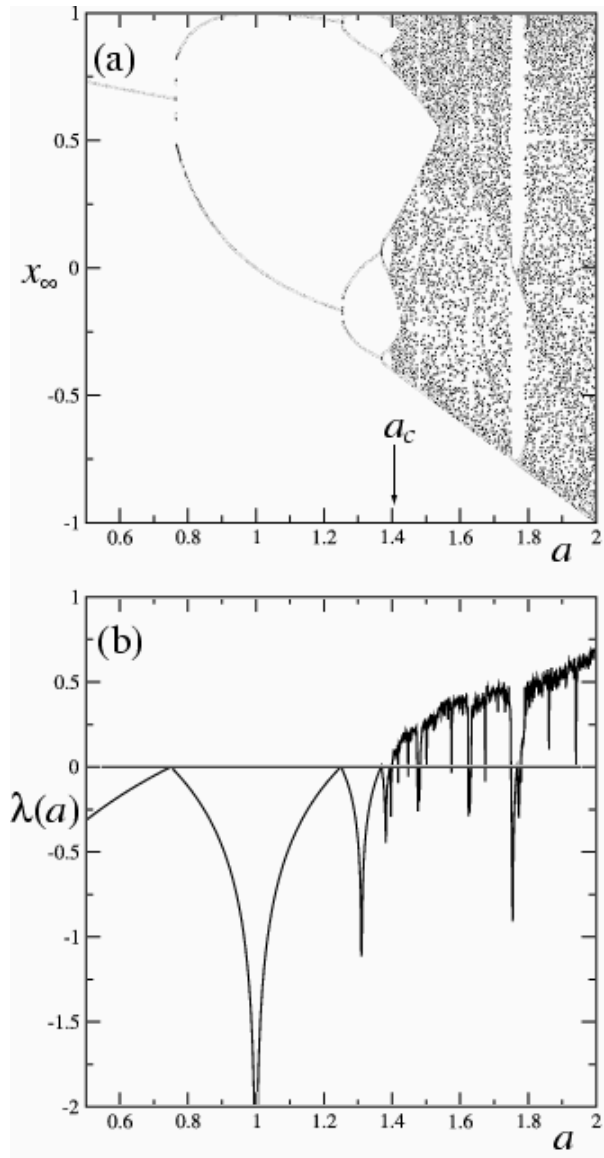}
\end{center}
\caption[]{
(a) Attractor of the logistic map ($z=2$) as a function of $a$.
The edge of chaos \index{edge of chaos} 
is at the critical value $a_c=  1.401155198...$
(b) Lyapunov exponent $\lambda$ as a function of $a$.}
\label{fig_lyap}
\end{figure}

\subsection{The logistic-like class of one-dimensional dissipative maps}
Although low-dimensional systems are often an
idealized representation of physical systems, they sometimes
offer the rare opportunity to obtain analytical results that
give profound insight in the comprehension of natural
phenomena.
This is certainly the case of the logistic-like class of
one-dimensional dissipative maps, that may be described by the
iteration rule
\begin{equation}
x_{t+1}=f(x_t)\equiv 1-a|x_t|^z,
\label{z-logistic}
\end{equation}
where $x\in[-1,1]$, $t=0,1,...$ is a discrete-time variable, 
$a\in[0,2]$ is a control parameter, and $z>1$
characterizes the universality class of the map (for $z=2$
one obtains one of the possible forms of the very well known
logistic map).
In particular, this class of maps captures the essential
mechanisms of chaos for dissipative systems, like the
period-doubling and the intermittency routes to chaos, and
constitute a prototypical example that is reported in
almost all textbooks in the area (see, e.g., \cite{beck,ott}).

As previously reported, the usual BG formalism applies to
situations where the dynamics of the system is sufficiently
chaotic, i.e., it is characterized by {\it positive} Lyapunov
coefficients. 
A central result in chaos theory, valid for several classes of
systems, is the identity between  
(the sum of positive) Lyapunov coefficients and  
the entropy growth (for Hamiltonian systems this result is called
Pesin theorem).
The failure of the classical BG
formalism and the possible validity of the nonextensive one
is related to the vanishing of the classical Lyapunov
coefficients $\lambda\equiv\lambda_1$ and to their replacement by
the generalized ones $\lambda_q$ (see Eq. (\ref{q-lyap})). 

Reminding that the sensitivity to initial conditions of
one-dimensional maps 
is associated to a single Lyapunov coefficient, 
the Lyapunov spectra of the logistic map ($z=2$), as a function 
of the parameter $a$, is displayed in Fig. \ref{fig_lyap},
together with the attractor 
$x_\infty=\left\{x\in[-1,1]:x=\lim_{t\to\infty}x_t\right\}$.
For $a$ smaller than a critical value $a_c=  1.401155198...$,
a zero Lyapunov coefficient is associated to the 
pitchfork bifurcations (period-doubling); while for $a>a_c$ 
the Lyapunov coefficient vanishes for example in correspondence
of the tangent bifurcations that generate the periodic windows
inside the chaotic region.   
In Ref. \cite{baldovinbifurcations}, using a 
renormalization-group (RG) analysis, it has been (exactly) proven that
the nonextensive formalism describes the dynamics 
associated to these critical points.
The sensitivity to initial conditions is in fact given
by the $q$-exponential Eq. (\ref{q-lyap}), 
with $q=5/3$ for pitchfork
bifurcations and $q=3/2$ for tangent bifurcations of any
nonlinearity $z$, while $\lambda_q$ depends on the order of the
bifurcation.
It is worthwhile to notice that these values are {\it not} deduced from
fitting; instead, they are analytically calculated by means of
the RG technique that describes the (universal) dynamics of these
critical points.

\begin{figure}
\begin{center}
\includegraphics[width=0.65\textwidth]{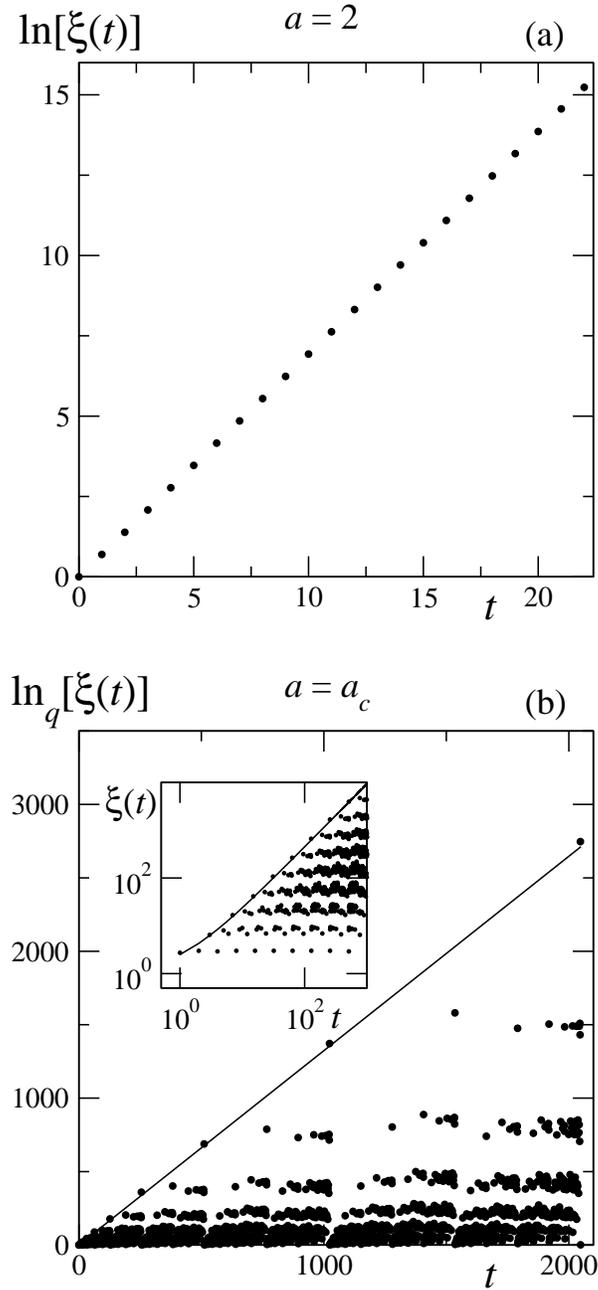}
\end{center}
\caption[]{Sensitivity to initial conditions for the logistic
map ($z=2$).
The dots represent $\xi(t)$ for
two initial data started at
$x_0=  1/2$ and $x_0^\prime\sim 1/2+10^{-8}$ (a),
and $x_0=  0$ and $x_0^\prime\sim 10^{-8}$ (b).
For $a=2$ the log-linear plot displays a linear increase with
a slope $\lambda=\ln 2$.
For $a=a_c$ the $q$-log-linear plot displays a linear increase 
of the upper bound (sequence $k=0$).
The solid line
is the function in Eq. (\protect\ref{q-lyap}), with
$q=  0.2445...$ and $\lambda_q= \ln \alpha_F / \ln 2=  1.3236...$. 
In the inset of (b) the same data represented in a log-log plot.}
\label{monster}
\end{figure}

Perhaps the most fascinating point of the logistic map 
is the {\it edge of chaos} $a=a_c$, that separates regular
behavior from chaoticity.
It is another point where 
the Lyapunov coefficient $\lambda$ vanishes, so
that no nontrivial information about the dynamics is attainable using the
classical approach. 
Nonetheless, once again the RG approach reveals to be extremely
powerful. 
Let us focus, for definiteness, on the case of the logistic map $z=2$.
Using the Feigenbaum-Coullet-Tresser RG transformation one can in fact show 
(see \cite{baldovinedge,pesinfulvioalberto} for details) that
the dynamics can be described by a series of subsequences
labelled by $k=0,1,...$,
characterized by the shifted iteration time 
\mbox{$t_k(n)=(2k+1)2^{n-k}-2k-1$}
($n$ is a natural number satisfying $n\geq k$), that are related
to the bifurcation mechanism.  
For each of these subsequences, the sensitivity to initial
conditions is given by the $q$-exponential Eq. (\ref{q-lyap}).
The value of $q$ (that is the same for all the subsequences) 
and $\lambda_q^{(k)}$ are deduced by one of the
Feigenbaum's universal constant $\alpha_F=2.50290...$ and are given
by 
\begin{equation}
q=1-\frac{\ln2}{\ln\alpha_F}=0.2445...\;\;\; 
{\rm and}\;\;\;
\lambda_q^{(k)}=\frac{\ln\alpha_F}{((2k+1)\ln2)}.
\end{equation}
In figure Fig. \ref{monster}(b) this function is drawn for the first
subsequence ($k=0$), together with the result of a numerical
simulation. 
For comparison purposes, Fig. \ref{monster}(a) shows that 
when the map is fully chaotic $\xi$ grows exponentially
with the iteration time, 
with the Lyapunov coefficient $\lambda=\ln2$ for $a=2$. 

\begin{figure}
\begin{center}
\includegraphics[width=.65\textwidth,angle=0]{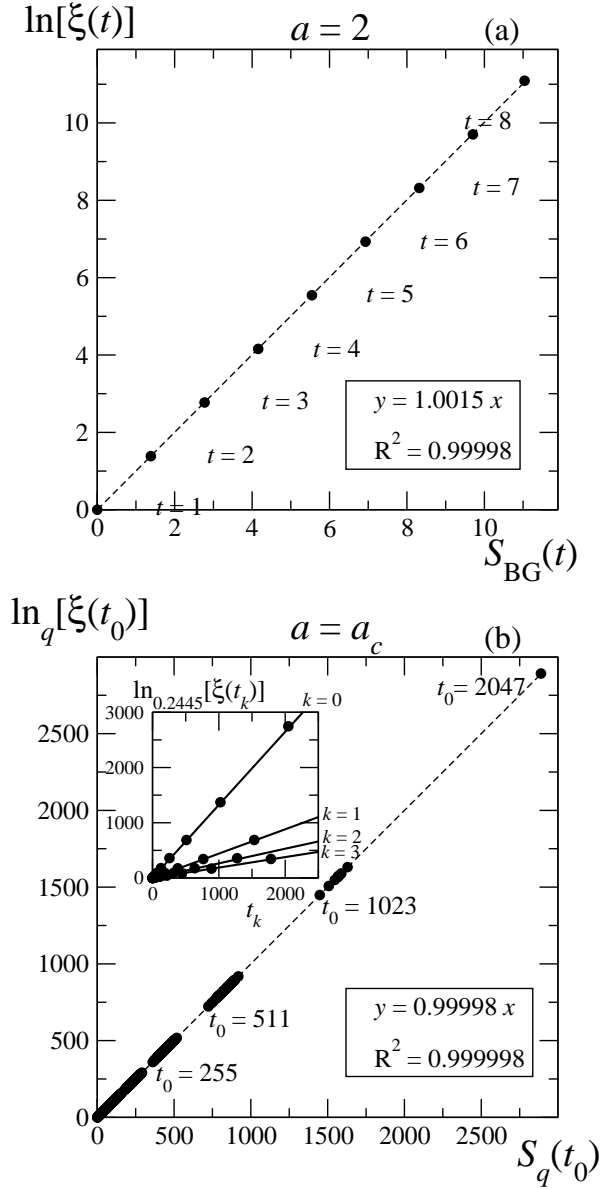}
\end{center}
\caption{ 
Numerical corroboration (full circles) of the
generalized Pesin identity $K_{q}^{(k)}=\lambda _{q}^{(k)}$ 
for the logistic map. 
On the vertical axis we plot the $q$-logarithm of $\xi$ (equal
to $\lambda _{q}^{(k)}t_k)$ and in the horizontal axis $S_{q} $
(equal to $K_{q}^{(k)}t_k$). 
Dashed lines are linear fittings.
(a) For $a=2$ the identity is obtained using the $BG$ formalism
$q=1$; 
while (b) at the edge of chaos $q=0.2445...$ must be
used. 
Numerical data in (b) are obtained partitioning the interval
$[-1,1]$ into cells of equal size $10^{-9}$ and considering a
uniform distribution of $10^5$ points inside the interval
$[0,10^{-9}]$ as initial ensemble; $\xi$ is calculated using, as inital conditions, the
extremal points of this same interval. 
A similar setup gives the numerical results in (a).
In the inset of (b) we plot the $q$-logarithm of $\xi$ 
as a function of the shifted time $t_k=(2k+1)2^{n-k}-2k-1$. 
Full lines are from the analytical result Eq. (\ref{q-lyap}). 
} 
\label{fig_pesin}
\end{figure}

For the edge of chaos it is also possible to proceed a step
further and consider the entropy production associated to an
ensemble of copies of the map, set 
initially out-of-equilibrium.
Remarkably enough, if (and only if) we consider the entropy $S_q$
precisely with $q=0.2445...$ for the definition of $K_q$ 
(see Eq. (\ref{kolmogorov})), for all the subsequences we obtain a {\it linear} dependance of $S_q$ with (shifted) time, i.e.,  
a generalized version of the Pesin identity:
\begin{equation}
K_q^{(k)}=\lambda_q^{(k)}.
\end{equation}
Fig. \ref{fig_pesin}(b) shows a numerical corroboration of this
analytical result. 
The $q$-logarithm of the sensitivity to initial conditions
plotted as a function of $S_q$ displays in fact a $45^\circ$
straight line for all iteration steps. 
Again, Fig. \ref{fig_pesin}(a) presents the analogous result
obtained for the chaotic situation $a=2$ using the $BG$ entropy. 
The inset of Fig. \ref{fig_pesin}(b) gives an insight of
Fig. \ref{monster}(b), showing that the linearity of the 
$q$-logarithm of $\xi$ with the iteration time is valid for all
the subsequences, once that the shifted time $t_k$ is used.

To conclude this illustration of low-dimensional nonextensivity, it is worthy to explicitly mention that the nontrivial value $q(z)$ (with $q(2)=0.2445...$) can be obtained from microscopic dynamics through at least {\it four} different procedures. These are: (i) from the {\it sensitivity to the initial conditions}, as lengthily exposed above and in \cite{tsallis97,baldovinedge,uriel}; (ii) from {\it multifractal geometry}, using 
\begin{equation}
\frac{1}{1-q(z)}=\frac{1}{\alpha_{min}(z)}- \frac{1}{\alpha_{max}(z)} =  \frac{(z-1) \ln \alpha_F(z)}{\ln 2}\;,
\end{equation}
whose details can be found in \cite{lyratsallis}; (iii) from {\it entropy production per unit time}, as exposed above and in \cite{pesincatania,pesinfulvioalberto};  and (iv) from {\it relaxation associated with the Lebesgue measure shrinking}, as can be seen in \cite{borgesetal} (see also \cite{lyrashrink}). 

\section{Final remarks}

Classical thermodynamics, valid for both classical and quantum systems, is essentially based on the following principles: $0^{th}$ principle ({\it transitivity of the concept of thermal equilibrium}), $1^{st}$ principle ({\it conservation of the energy}), $2^{nd}$ principle ({\it macroscopic irreversibility}), $3^{rd}$ principle ({\it vanishing entropy at vanishing temperature}), and $4^{th}$ principle ({\it reciprocity of the linear nonequilibrium coefficients}). All these principles are since long known to be satisfied by Boltzmann-Gibbs statistical mechanics. However, a natural question arises: {\it Is BG statistical mechanics the only one capable of satisfying these basic principles?} The answer is {\it no}. Indeed, the present nonextensive statistical mechanics appears to also satisfy all these five principles (thermal equilibrium being generalized into stationary or quasistationary state or, generally speaking, {\it metaequilibrium}), as we have argued along the present review. The second principle in particular has received very recently a new confirmation \cite{second}.    

The connections between the BG entropy and the BG exponential energy distribution are since long established through various standpoints, namely steepest descent, large numbers, microcanonical counting and variational principle. The corresponding $q$-generalization is equally available in the literature nowadays. Indeed, through {\it all these procedures}, the entropy $S_q$ has been connected to the $q$-exponential energy distribution, in particular in a series of works by Abe and Rajagopal (see \cite{tsallisinswinney, tsallisingellmann} and references therein).    

In addition to all this, $S_q$ shares with the BG entropy concavity, stability, finiteness of the entropy production per unit time. Other well known entropies, such as the Renyi one for instance, do not.

Summarizing, the dynamical scenario which emerges is that whenever ergodicity (or at least an ergodic sea) is present, one expects the BG concepts to be the adequate ones. But when ergodicity fails, very particularly when it does so in a special hierarchical (possibly multifractal) manner, one might expect the present nonextensive concepts to naturally take place. Furthermore, we conjecture that, in such cases, the visitation of phase space occurs through some kind of scale-free topology.\\

{\bf ACKNOWLEDGMENTS:}

The present effort has benefited from partial financial support by CNPq, Pronex/MCT, Capes and Faperj (Brazilian agencies) and SIF (Italy).


\newpage

\begin{thebibliography}{99}      
 
\bibitem {lesche}B. Lesche, J. Stat. Phys. \textbf{27}, 419 (1982).                                                                                      %

\bibitem {einstein}A. Einstein, Annalen der Physik \textbf{33}, 1275 (1910) 
[Translation:\ A. Pais, {\it Subtle is the Lord...} (Oxford University Press, 1982)].

\bibitem {fermi}E. Fermi, {\it Thermodynamics} (1936).

\bibitem{tisza}L. Tisza, Annals Phys. {\bf 13}, 1 (1961) [or in 
{\it Generalized thermodynamics}, (MIT Press, Cambridge, 1966), p. 
123]. 

\bibitem{landsberg}P.T. Landsberg, {\it Thermodynamics and 
Statistical Mechanics}, (Oxford University Press, Oxford, 1978; also 
Dover, 1990), page 102.  

\bibitem {tsallis}C. Tsallis, J. Stat. Phys. \textbf{52} (1988), 479.

\bibitem{takens}F. Takens, in {\it Structures in dynamics - Finite 
dimensional deterministic studies}, eds. H.W. Broer, F. Dumortier, S.J. 
van Strien and F. Takens (North-Holland, Amsterdam, 1991), page 253. 

\bibitem {shannon}C.E. Shannon, and W. Weaver, The mathematical theory of
communication Urbana University of Illinois Press, Urbana, 1962.

\bibitem {khinchin}A: I. Khinchin, Mathematical foundations of informations
theory, Dover, New York, 1957.

\bibitem {santos}R. J. V. Santos, J. Math. Phys. \textbf{38}, 4104 (1997).

\bibitem {abe}S. Abe, Phys. Lett. A, \textbf{271}, 74 (2000).

\bibitem{plastinosCondMat}A.R. Plastino and A. Plastino, {\it Condensed Matter Theories}, ed. E. Ludena, {\bf 11}, 327 (Nova Science Publishers, New York, 1996).

\bibitem{plastinosBJP}A. Plastino and A.R. Plastino, in {\it Nonextensive Statistical Mechanics and Thermodynamics}, eds. S.R.A. Salinas and C. Tsallis, Braz. J. Phys. {\bf 29}, 50 (1999).

\bibitem {abe2}S. Abe, Phys. Lett. A, \textbf{224}, 326 (1997).

\bibitem {abe3}S. Abe, Phys. Rev. E, \textbf{66}, 046134 (2002).

\bibitem {renyi}A. Renyi, Proc. 4$^{th}$ Berkeley Symposium (1960).

\bibitem {lv}P. T. Landsberg and V. Vedral, Phys. Lett. A, \textbf{247}, 211 (1998).

\bibitem {ra}A. K. Rajagopal and S. Abe, Phys. Rev.\ Lett. \textbf{83}, 1711 (1999).

\bibitem {tsallis98}C. Tsallis, R. S. Mendes and A. R. Plastino, Physica A \textbf{261} (1998), 534.

\bibitem{brigatti}C. Tsallis and E. Brigatti, to appear in {\it Extensive and non-extensive entropy and statistical mechanics}, special issue of Continuum Mechanics and Thermodynamics, ed. M. Sugiyama (Springer, 2003), in press [cond-mat/0305606].

\bibitem{curado}E.M.F. Curado and C. Tsallis, J. Phys. A {\bf 24}, L69 
(1991) [Corrigenda: {\bf 24}, 3187 (1991) and {\bf 25}, 1019 (1992)].

\bibitem{tsallisinswinney}C. Tsallis, to appear in a special volume of Physica D entitled {\it Anomalous Distributions, Nonlinear Dynamics and Nonextensivity}, eds. H.L. Swinney and C. Tsallis (2003), in preparation.

\bibitem {ra2}A. K. Rajagopal, Phys. Rev.\ Lett. \textbf{76}, 3469 (1996).

\bibitem {tsallis97}C. Tsallis, A. R. Plastino and W. M. Zheng, Chaos, Solitons and
fractals \textbf{8} (1997), 885.

\bibitem {mariz}A. M. Mariz, Phys. Lett. A \textbf{165}, 409 (1992).

\bibitem {ramshaw}J. D. Ramshaw, Phys. Lett. A \textbf{175}, 169 (1993).

\bibitem{secondprinciple}S. Abe and A.K. Rajagopal, Phys. Rev. Lett. (2003), in press [cond-mat/0304066]. 

\bibitem {plastino93}A. R. Plastino, A. Plastino, Phys. Lett. A \textbf{177}, 384 (1993).

\bibitem{cacerestsallis}M.O. Caceres and C. Tsallis, private discussion (1993).

\bibitem {caceres}M. O. Caceres, Physica A \textbf{218}, 471 (1995).

\bibitem {chame}A. Chame and V. M. De Mello, Phys. Lett. A \textbf{228}, 159 (1997).

\bibitem {chame2}A. Chame and V. M. De Mello, J. Phys. A \textbf{27}, 3663 (1994).

\bibitem{tsallis95}C. Tsallis, Chaos, Solitons and Fractals \textbf{6}, 539 (1995).

\bibitem{pesincatania}V. Latora, M. Baranger, A. Rapisarda and C. Tsallis, Phys. Lett. A {\bf 273}, 97 (2000).

\bibitem{pesinfulvioalberto}F. Baldovin and A. Robledo, cond-mat/0304410. 

\bibitem{tsallisingellmann}C. Tsallis, in {\it Nonextensive Entropy: Interdisciplinary Applications}, eds. M. Gell-Mann and C. Tsallis (Oxford University Press, 2003), to appear.

\bibitem{biblio}http://tsallis.cat.cbpf.br/biblio.htm

\bibitem{antoniruffo}M. Antoni and S. Ruffo, Phys. Rev. E {\bf 52}, 2361 (1995).

\bibitem{fisheretal}M.E. Fisher, Arch. Rat. Mech. Anal. {\bf 17}, 377 (1964), J. Chem. Phys. {\bf 42}, 3852 (1965) and J. Math. Phys. {\bf 6}, 1643 (1965); M.E. Fisher and D. Ruelle, J. Math. Phys. {\bf 7}, 260 (1966); M.E. Fisher and J.L. Lebowitz, Commun. Math. Phys. {\bf 19}, 251 (1970).  

\bibitem{celiaconstantino}C. Anteneodo and C. Tsallis, Phys. Rev. Lett. {\bf 80}, 5313 (1998).

\bibitem{giansanti}A. Campa, A. Giansanti, D. Moroni and C. Tsallis, Phys. Lett. A {\bf 286}, 251 (2001).

\bibitem{vitoandreaconstantino}V. Latora, A. Rapisarda and C. Tsallis, Phys. Rev. E {\bf 64}, 056134 (2001).

\bibitem{aging}M.A. Montemurro, F. Tamarit and C. Anteneodo, Phys. Rev. E {\bf 67}, 031106 (2003).

\bibitem{cabral}B.J.C. Cabral and C. Tsallis, Phys. Rev. E {\bf 66}, 065101(R) (2002).

\bibitem{tsallisreply}C. Tsallis, cond-mat/0304696 (2003).

\bibitem{moyano}L.G. Moyano, F. Baldovin and C. Tsallis, cond-mat/0305091. 

\bibitem{physicaa}C.Tsallis, E.P.Borges, F.Baldovin, Physica A {\bf 305}, 1 (2002). 

\bibitem{nobre}F.D. Nobre and C. Tsallis, Phys. Rev. E (2003), in press [cond-mat/0301492].

\bibitem{ernestoising}E.P. Borges, C. Tsallis, A. Giansanti and D. Moroni, to appear in a volume honoring S.R.A. Salinas (2003) [in Portuguese].

\bibitem{doye}J.P.K. Doye, Phys. Rev. Lett. {\bf 88}, 238701 (2002).

\bibitem{second}S. Abe and A.K. Rajagopal, Phys. Rev. Lett. (2003), in press [cond-mat/0304066]. 

\bibitem{beck}  
C. Beck and F. Schlogl, {\it Thermodynamics
of Chaotic Systems} (Cambridge University Press, UK, 1993).

\bibitem{ott}
E. Ott, {\it Chaos in dynamical systems} (Cambridge
University Press, UK, 1993).

\bibitem{baldovinbifurcations}  
F. Baldovin and A. Robledo, Europhys. Lett. {\bf 60}, 518 (2002).

\bibitem{baldovinedge} 
F. Baldovin and A. Robledo, Phys. Rev. E 66, 045104(R) (2002).

\bibitem{uriel}U.M.S. Costa, M.L. Lyra, A.R. Plastino and C. Tsallis, Phys. Rev. E {\bf 56}, 245 (1997).

\bibitem{lyratsallis}M.L. Lyra and C. Tsallis, Phys. Rev. Lett. {\bf 80}, 53 (1998); M.L. Lyra, Ann. Rev. Comp. Phys. , ed. D. Stauffer (World Scientific, Singapore, 1998), page 31.

\bibitem{borgesetal}E.P. Borges, C. Tsallis, G.F.J. Ananos and P.M.C. Oliveira,  Phys. Rev. Lett. {\bf 89}, 254103 (2002); see also Y.S. Weinstein, S. Lloyd and C. Tsallis, Phys. Rev. Lett. {\bf 89}, 214101 (2002) for a quantum illustration.

\bibitem{lyrashrink}F.A.B.F. de Moura, U. Tirnakli and M.L. Lyra, Phys. Rev. E {\bf 62}, 6361 (2000).

\end{thebibliography}
\end{document}